\documentclass[12pt]{article}
\begin{document}

\begin{center}
{\Large {\bf COMMENT on Yuan {\it et al}. }}

\vspace*{0.15in}

Fan Wang

\vspace*{-0.11in}

Center for Theoretical Physics and Department of Physics, \\
Nanjing University, Nanjing, 210008, P. R. China\\

\vspace*{0.15in}

T. Goldman

\vspace*{-0.11in}

Theoretical Division, Los Alamos National Laboratory\\
Los Alamos, NM 87545, USA
\end{center}

\begin{flushright}
\vspace{-3.2in}
{LA-UR-00-51}\\
{nucl-th/0001019}\\
\vspace{3.4in}
\end{flushright}

\vspace*{-0.7in}

It has come to our attention that the work of Yuan et
al.$^{\cite{yetal}}$ is being represented as definitive regarding the
existence of a particular nonstrange, isoscalar, spin three dibaryon,
$d^{*}$, which we$^{\cite{rpprs}}$, among
others$^{\cite{mulders,yazaki,tetal,KF,CWW}}$, have proposed exists. We
therefore wish to comment on this paper and its relation to ours and
those of others.

In our work$^{\cite{rpprs}}$, the $d^{*}$ dibaryon is constructed in
terms of its constituent quark wavefunctions, which include significant
distortion from those found in isolated baryons. It should be noted
that these wavefunctions do not collapse into a simple, spherical
system as proposed, for example, by Jaffe$^{\cite{JH}}$ for the $H$
dibaryon. Our techniques have allowed us to demonstrate that this
picture adequately reproduces known low energy baryon-baryon scattering
amplitudes with only one or two fitting parameters.

At the other extreme are theories of baryon interactions without any
quark substructure whatsoever. These also, in terms of effective
potentials or meson exchanges, adequately reproduce known low energy
baryon-baryon scattering amplitudes, albeit with a multitude of fitting
parameters. Such approaches tend not to predict deeply bound dibaryons,
with the exception of Ref.\cite{KF}.  Rather, the states examined are
generally found to have a character not very dissimilar from the
deuteron, with small binding energies predicted$^{\cite{nisk}}$, if
any.

Approaches such as that of Yuan et al.$^{\cite{yetal}}$ and similarly
those of Thomas et al.$^{\cite{tetal}}$ or Wilets et
al.$^{\cite{wetal}}$ take an intermediate view, including quark
substructure of distinct baryons, but variously restricting
interactions between quarks as due to exchanges of mesons or other
collective fields, exchanges of gluons, or both. Some change of
internal baryon structure is allowed in these models. Interestingly,
the approaches that include or are restricted to only meson exchanges
may be characterized as generally predicting dibaryon states with
binding energies intermediate between the two approaches referred to
above. They also can well describe low energy baryon-baryon scattering
amplitudes or nuclei, albeit again with a significant number of
parameters which must be fit to data.

We wish to suggest that the proper scientific response to this range of
models and results is not to make a theoretical judgment about which is
more likely to be correct. Rather we feel that it behooves us all to
acknowledge that our understanding is limited and to recognize two
issues: One is the question of the range of scales over which nucleons
and mesons may be viewed as having rigid internal structures
uninfluenced by their surroundings. The second is the question of
whether or not quark propagation, at least at low energies, can be
coherent over ranges larger than about one fermi, or conversely,
whether their propagation must always be re-expressed in terms of
composite, colorless, degrees of freedom.

Yuan et al. also criticize the manner in which we use a nonrelativistic
form of the confinement potential. However, as we have explained in
Ref.\cite{pwg}, our nonrelativistic model Hamiltonian is an extended,
effective matrix element approach rather than simply a potential model.
It is sufficient to define a quantum mechanical model if all of the
matrix elements of the Hamiltonian in the model Hilbert space have been
fixed. Their concern regarding the nonorthogonality of the left and
right centered orbits in our approach is misplaced since the fixing of
the matrix elements by our model assumption is clearly defined. As has
been emphasized before, our understanding of confinement is limited so
it is perhaps unwise to narrowly restrict model approaches to any
description of confinement.

Our view is that it is indeed fortunate that this range of models with
differing physics assumptions produces a range of predicted masses for
the $d^{*}$. This allows the experimental search for such a state, and
the mass determination if it is observed, to distinguish among pictures
which a priori have similar strengths for their claims of accuracy.
This makes experiments, such as the proton induced excitation discussed 
by Wong$^{\cite{CWW}}$ and possible experiments involving electron
scattering$^{\cite{QSW}}$, extremely important to carry out. Whatever
the result of these experiments, they will have much to say about how
we should view the realm of low energy strong interactions in terms of
quarks.

This research is supported in part by the National Science Foundation 
of China and in part by the Department of Energy under contract
W-7405-ENG-36.

\pagebreak

\end{document}